\documentclass[aps,12pt,showpacs]{revtex4}%
\usepackage{amsfonts}
\usepackage{amsmath}
\usepackage{amssymb}
\usepackage{graphicx}%
\providecommand{\U}[1]{\protect\rule{.1in}{.1in}}

\begin{document}
\title{Fractional Boundaries for Fluid Spheres }
\author{S. Bayin$^{1}$, E.N. Glass$^{2}$, J.P. Krisch$^{2}$}
\affiliation{$^{1}$Physics Department, Middle East Technical University, Ankara, Turkey}
\author{$^{2}$\textit{Physics Department, University of Michigan, Ann Arbor, MI}}
\date{23 October 2005}

\begin{abstract}
An single Israel layer can be created when two metrics adjoin with no
continuous metric derivative across the boundary.\ The properties of the layer
depend only on the two metrics it separates. By using a fractional derivative
match, a family of Israel layers can be created between the same two
metrics.\ The family is indexed by the order of the fractional
derivative.\ The method is applied to Tolman IV and V interiors and a
Schwarzschild vacuum exterior.\ The method creates new ranges of modeling
parameters for fluid spheres. A thin shell analysis clarifies pressure/tension
in the family of boundary layers.

\end{abstract}

\pacs{\ 04.20. -q, 04.20.Jb, 04.20.Dg}
\maketitle

\section{Introduction}

There is long standing interest in fluid sphere solutions, largely because of
their astrophysical implications. An astrophysical model is often an interior
fluid sphere metric matched to a Schwarzschild vacuum or Kottler exterior
across a bounding surface.\ The standard technique matches metric functions
and extrinsic curvatures on the boundary. When the extrinsic curvatures do not
match, an Israel boundary layer \cite{Isr77},\cite{Poi04} can be created.\ The
layer depends only the properties of the two bounding metrics.\ Methods that
will create a family of surface layers between the bounds could prove useful
in exploring models of spheres with variable crusts. One way of creating
variable surface layers is to modify the boundary conditions at the
fluid-vacuum interface. \ 

While an extrinsic curvature match is the boundary condition currently most
used, there are three types of boundary conditions that have been used to
match analytic solutions across non-null boundaries. The three methods have
been discussed by Bonnor and Vickers \cite{BV81}, and they all involve
derivatives of the metric functions.\ The boundary conditions can be
generalized by broadening the idea of derivatives to include fractional
derivatives \cite{OS74},\cite{SKB02}.\ 

There are two simple ways to proceed with the generalization. The first is to
assume a straight fractional derivative match on the boundary metrics and then
to use the fractional relations in the usual formalism for the boundary
stress-energy. This would be a generalization of the Lichnerowicz boundary
condition. It would not generalize the extrinsic curvature to fractional
values. The second would be to use fractional derivatives to define a
fractional extrinsic curvature and then use it to define a fractional boundary
layer.\ This would be a generalization of the usual Lie derivative to
fractional values. The use of fractional calculus is motivated by the possible
fractional nature of the growth processes forming the boundary layer.
Fractional transport processes are one of the main areas of application for
fractional calculus, and boundary layers formed by these processes could
reflect this fractional formation process. \textbf{\ }

Beyond the fractional generalizations of techniques and tensor functions, one
must consider the various definitions of fractional differentiation.\ Use of
fractional calculus in diverse areas of physics has increased enormously since
fractional derivatives were first considered by Leibnitz and L'Hospital
\cite{OS74} in 1695. Many different definitions have been proposed for
different applications. In this article we use the Caputo form of
the\ Riemann-Liouville and Weyl definitions. The Caputo derivative is an
integral transform of the regular partial derivative and preserves zero
fractional derivative of a constant.\ While considering generalizations of
relativistic gravity to include fractional calculus, the different definitions
must be explored to determine their applicability.\textbf{ \ }

This work has two goals: first to develop a variable layer model that could be
applied to astrophysical problems, and second to better understand the role
that fractional derivatives might play within a general relativistic
framework. In this article we will apply the first method and use fractional
derivatives to create a family of Israel boundary layers between two bounding
metrics.\ The family is parameterized by the order of the fractional
derivative and may be used to model fluid spheres with variable crusts.\ Even
when a regular derivative match is possible, the fractional match will broaden
the parameter ranges for the fluid interior.

In the next section we discuss the metrics and describe the boundary layer. In
the third section several models are considered: the Misner-Zapolsky (MZ)
solution \cite{MZ64a},\cite{MZ64b}, and Tolman's solutions IV and V
\cite{Tol39}. The thin shell pressure balance is treated in the fourth section
and some details of the fractional match are discussed in section
five.\ Details of the fractional derivatives and the standard fluid sphere
formalism are given in Appendices. \ 

\section{Theoretical Framework}

\subsection{The spacetime}

The two regions to be considered are covered by an exterior Schwarzschild
solution bounding an interior spherical fluid.\ The metrics are, with
functions $\psi_{\text{Sch}}=1-2m_{0}/y,$ $\nu(r),$ $\lambda(r),$ $H(r)$%
\begin{align}
\text{Exterior}  &  \text{: }g_{\alpha\beta}^{\text{Sch}}dx^{\alpha}dx^{\beta
}=-\psi_{\text{Sch}}dt^{2}+\psi_{\text{Sch}}^{-1}dy^{2}+y^{2}d\Omega^{2}\\
\text{Interior}  &  \text{:\ }g_{\alpha\beta}^{\text{fluid}}dx^{\alpha
}dx^{\beta}=-e^{\nu}d\tau^{2}+e^{\lambda}dr^{2}+H^{2}d\Omega^{2}%
\end{align}
The bounding surface is located at $y=y_{0}$ in the exterior and $r=R_{0}$ in
the interior.\ The corresponding normals to the surface are
\begin{align}
\text{Exterior}  &  \text{: }n_{\mu}^{\text{E}}dx^{\mu}=\psi_{\text{Sch}%
}^{-1/2}dy\\
\text{Interior}  &  \text{: }n_{\mu}^{\text{I}}dx^{\mu}=e^{\lambda/2}dr
\end{align}
Fractional derivatives leave these metrics unchanged. Our fractional extension
provides a crust layer between the interior and exterior metrics.

\subsection{Matching Conditions}

On the boundary, the metric match conditions are%
\begin{align}
(1-2m_{0}/R_{0})  &  =e^{\nu(R_{0})}\\
R_{0}  &  =H(R_{0}).\nonumber
\end{align}
The second matching condition is the extrinsic curvature match, $K_{b}^{a}$,
on the bounding surface. If the curvatures do not match, an Israel boundary
layer is created.\ The stress-energy content of the Israel layer is
constructed from the mismatch in the extrinsic curvatures. The stress-energy
of the boundary layer is \cite{Poi04}
\begin{equation}
-8\pi S_{b}^{a}=\text{ }<K_{b}^{a}>-<K>g_{b}^{a}.
\end{equation}
Here $K=K_{a}^{a}$. The stress-energy components on the boundary are:
\begin{align*}
-8\pi S_{0}^{0}  &  =[<K_{0}^{0}>-<K_{0}^{0}+2K_{\theta}^{\theta}>g_{0}%
^{0}]=-2<K_{\theta}^{\theta}>\\
-8\pi S_{\theta}^{\theta}  &  =-8\pi S_{\phi}^{\phi}=[<K_{\theta}^{\theta
}>-<K_{0}^{0}+2K_{\theta}^{\theta}>g_{\theta}^{\theta}]=-[<K_{\theta}^{\theta
}>+<K_{0}^{0}>]
\end{align*}
and the stress-energy of the boundary is%
\begin{align}
8\pi S_{0}^{0}  &  =\frac{1}{(g_{yy}^{E})^{1/2}}\frac{g_{\theta\theta,y}^{E}%
}{g_{\theta\theta}^{E}}-\frac{1}{(g_{rr}^{I})^{1/2}}\frac{g_{\theta\theta
,r}^{I}}{g_{\theta\theta}^{I}}\label{stress-energy}\\
8\pi S_{\theta}^{\theta}  &  =8\pi S_{\phi}^{\phi}=\frac{1}{2}\left[  \frac
{1}{(g_{yy}^{E})^{1/2}}\frac{g_{\theta\theta,y}^{E}}{g_{\theta\theta}^{E}%
}-\frac{1}{(g_{rr}^{I})^{1/2}}\frac{g_{\theta\theta,r}^{I}}{g_{\theta\theta
}^{I}}\right]  +\frac{1}{2}\left[  \frac{1}{(g_{yy}^{E})^{1/2}}\frac
{g_{00,y}^{E}}{g_{00}^{E}}-\frac{1}{(g_{rr}^{I})^{1/2}}\frac{g_{00,r}^{I}%
}{g_{00}^{I}}\right]  .\nonumber
\end{align}

\subsection{Match of fractional derivatives}

The stress-energy of the Israel layer is evaluated on the boundary between the
interior and exterior metrics. The actual finite thickness boundary layer is
modeled by the single bounding surface at $r=R_{0}$.\ The stress-energy
content is governed by regular derivatives of the metric functions.\ The
metric match coupled with some derivative match of the metric on the layer,
sets relations between the parameters of the interior and exterior solutions.
With the usual extrinsic curvature or other derivative matches, the properties
of the layer are set by the parameters of the bounding metric.\ With a
fractional match, the order of the fractional derivative enters along with the
other parameters and a family of fractional boundary layers is created. The
fluid sphere examples considered in this paper have boundary metrics of the
form%
\[
ds^{2}=-F(r)dt^{2}+r^{2}d\Omega^{2}.
\]
The fractional match is applied only to the differing part of the Israel layer
metric, the $g_{00}$ metric potential. The actual calculation of the
fractional derivatives involves a choice of definition. We use the Caputo
definition (see Appendix A) with the $(0\leq r\leq R_{0})$ Riemann-Liouville
limits for the interior and the $(R_{0}\leq r\leq\infty)$ Weyl limits for the
exterior. The limits themselves, as well as the choice of different limits for
interior and exterior derivatives reflect the non-locality of the fractional
derivative operation. Non-locality in fractional time derivatives is an
expression of system memory \cite{Hil00}.\ It has proven especially useful in
modeling jump processes with long wait times \cite{Mai97}.\ Similarly, spatial
non-locality\ implies the derivative on the boundary depends on values away
from the boundary; fractional spatial derivatives have been useful in modeling
processes with very large jump distances \cite{MBB99}. When the jump distance
depends on the jump time, fractional time and spatial derivatives enter into
the transport equations \cite{BMM04}.\ The examples discussed here are static
but the structure of the boundary layer could reflect the transport process.
The fractional matching condition is%
\[
\frac{1}{\Gamma(n-\alpha)}%
{\displaystyle\int\limits_{0}^{R_{0}}}
\frac{d^{n}F_{I}(x)/dx^{n}}{(r-x)^{\alpha-n+1}}dx=\frac{(-1)^{n-1}}%
{\Gamma(n-\alpha)}%
{\displaystyle\int\limits_{R_{0}}^{\infty}}
\frac{d^{n}F_{E}(x)/dx^{n}}{(x-r)^{\alpha-n}}dx
\]
and is applied at $r=R_{0}$. We note that the single layer at $r=R_{0}$ only
approximates a boundary of finite thickness and that using a non-local
operator might be a better approximation to the actual match over a finite
thickness than the usual derivative match over a zero thickness surface.

In the next sections, we apply the formalism to Tolman IV and V solutions.

\section{Model Calculations}

\subsection{Tolman's Solution V}

\subsubsection{The solution}

We consider a parametrization of Tolman's Vth solution \cite{Tol39}%
,\cite{IS82}.\ The metric, with constants $n$ and $C$, is%
\begin{equation}
ds^{2}=-(r/r_{0})^{N_{1}}dt^{2}+a(1-aCr^{2+b})^{-1}dr^{2}+r^{2}d\Omega^{2}.
\end{equation}
The parameters formed from $n$ are%
\begin{align*}
N_{1}  &  =4n/(1+n),\text{ \ }N_{2}=1+6n+n^{2},\\
a  &  =\frac{N_{2}}{(1+n)^{2}},\text{ \ \ }b=\frac{N_{1}(1-n)}{(1+3n)}.
\end{align*}
The interior density and pressure for this solution are
\begin{align}
8\pi\rho &  =(\frac{4n}{N_{2}})\frac{1}{r^{2}}+C(3+b)r^{b},\\
8\pi P  &  =(\frac{4n^{2}}{N_{2}})\frac{1}{r^{2}}-C\frac{(1+5n)}{1+n}%
r^{b}.\nonumber
\end{align}

For $C=0$,$\ $ the solution reduces to the MZ solution \cite{MZ64a}%
,\cite{MZ64b},\cite{CB77}. This solution was originally used to describe
neutron star models with equation of state $P=n\rho$. The solution with
$C$~$=$~$0\ $does not admit a zero-pressure boundary, the $C$~$\neq$%
~$0\ $solution does.\ Both solutions are singular at the origin and are used
to represent an ultrahigh density core. The MZ solution, lacking a vacuum
boundary, is generally matched to a gaseous envelope. It may be more realistic
in some cases to match these solutions to a crust with surface stresses.

For $C$~$\neq$~$0$, the zero-pressure boundary, $R_{z}$, relates constants $C$
and $n$\
\begin{equation}
C=\frac{4n^{2}(1+n)}{(1+5n)N_{2}}\frac{1}{R_{z}^{2+b}}. \label{c-n-eqn}%
\end{equation}
Substituting for $C$, the pressure and density can be written as%
\begin{align}
8\pi P_{C\neq0}  &  =\frac{4n^{2}}{N_{2}}\left(  \frac{1}{r^{2}}-\frac{r^{b}%
}{R_{z}^{2+b}}\right) \label{p-c}\\
8\pi\rho_{C\neq0}  &  =\frac{4n}{N_{2}}\left(  \frac{1}{r^{2}}\right)
+\frac{4n^{2}}{N_{2}}\frac{(n+3)}{(1+3n)}\left(  \frac{r^{b}}{R_{z}^{2+b}%
}\right)  . \label{rho-c}%
\end{align}
The condition $P_{C\neq0}\geq0$ requires%
\begin{equation}
R_{0}\leq R_{z}%
\end{equation}

Fractional matching will allow a broader family of sphere sizes. Below, we
graph values for the case $n=1/3$. For $C=0$, there is no zero-pressure
boundary and no constraint.

\subsubsection{Matching Conditions}

The matching conditions are the same for any $C$ value. Matching the interior
metric to vacuum Schwarzschild we find
\[
1-2m_{0}/R_{0}=(R_{0}/r_{0})^{N_{1}}.
\]
[recall $N_{1}=4n/(1+n)$]
\[
\frac{2m_{0}}{R_{0}^{1+\alpha}}~\Gamma(1+\alpha)=(\frac{R_{0}}{r_{0}})^{N_{1}%
}(\frac{N_{1}}{R_{0}^{\alpha}})\frac{\Gamma(N_{1})}{\Gamma(N_{1}+1-\alpha)}.
\]
Combining the two relations we find the scaled radius of the interior is
\begin{equation}
\frac{R_{0}}{2m_{0}}=1+\frac{\Gamma(1+\alpha)\Gamma(N_{1}+1-\alpha)}%
{\Gamma(1+N_{1})} \label{frac-match-N}%
\end{equation}
Note that the boundary radius is always greater than $2m_{0}.$ For $\alpha<1,$
there are no limits imposed by Eq.(\ref{frac-match-N}).\ For $\alpha\geq1$ we
require%
\[
1+N_{1}>\alpha
\]
The metric parameter $r_{0}$ is described by
\begin{equation}
(\frac{r_{0}}{R_{0}})^{N_{1}}=1+\frac{\Gamma(1+N_{1})}{\Gamma(1+\alpha
)\Gamma(N_{1}+1-\alpha)}%
\end{equation}
The sizes of the fractional spheres are discussed in section V. \ 

\subsubsection{The Crust Stress-Energy}

The stress-energy of the crust for general $C$ is, with $\gamma_{0}%
:=\sqrt{1-2m_{0}/R_{0}}$%
\begin{align*}
8\pi S_{0}^{0}  &  =\frac{2}{R_{0}}\left[  \gamma_{0}-(1+n)N_{2}^{-1/2}%
\sqrt{1-aCR_{0}^{2+b}}\right] \\
8\pi S_{\theta}^{\theta}  &  =8\pi S_{\phi}^{\phi}=\frac{1}{2R_{0}}\left[
\gamma_{0}+1/\gamma_{0}-(1+n)(2+N_{1})N_{2}^{-1/2}\sqrt{1-aCR_{0}^{2+b}%
}\right]
\end{align*}
For $C\neq0,$ the boundary layer has a stress-energy content, [recall
$N_{1}=4n/(1+n)$, $N_{2}=1+6n+n^{2}$]%
\begin{align}
8\pi S_{0}^{0}  &  =\frac{2}{R_{0}}\left[  \gamma_{0}-(1+n)N_{2}^{-1/2}%
\sqrt{1-\frac{nN_{1}}{(1+5n)}(R_{0}/R_{z})^{2+b}}\right] \\
8\pi S_{\theta}^{\theta}  &  =8\pi S_{\phi}^{\phi}=\frac{1}{2R_{0}}\left[
\gamma_{0}+1/\gamma_{0}-2(1+3n)N_{2}^{-1/2}\sqrt{1-\frac{nN_{1}}{(1+5n)}%
(R_{0}/R_{z})^{2+b}}\right] \nonumber
\end{align}
For $C=0$ the fluid energy density and stress are \
\begin{align}
8\pi S_{0}^{0}  &  =(2/R_{0})\left[  \gamma_{0}-(1+n)N_{2}^{-1/2}\right] \\
&  =(2/R_{0})\left[  (R_{0}/r_{0})^{2n/(1+n)}-(1+n)N_{2}^{-1/2}\right]
\nonumber\\
8\pi S_{\theta}^{\theta}  &  =8\pi S_{\phi}^{\phi}=(1/R_{0})\left[
(1-m_{0}/R_{0})/\gamma_{0}-(1+3n)N_{2}^{-1/2}\right] \nonumber
\end{align}
and describe a much richer modeling environment. \ 

\subsection{Tolman's Solution IV}

\subsubsection{Metric and Stress-Energy}

This solution describes an object with finite central presssure and density. A
stiff fluid core is not possible in this model. The interior metric for this
solution is, with constants $A$, $B$, and $C$%
\begin{equation}
ds^{2}=-B^{2}(1+r^{2}/A^{2})dt^{2}+\frac{1+2r^{2}/A^{2}}{(1-r^{2}%
/C^{2})(1+r^{2}/A^{2})}dr^{2}+r^{2}d\Omega^{2}. \label{tol-iv-met}%
\end{equation}
The interior density and pressure are%
\begin{align}
8\pi\rho &  =\frac{1}{A^{2}}\left[  \frac{1+3(A^{2}/C^{2}+r^{2}/C^{2}%
)}{1+2r^{2}/A^{2}}+2\frac{1-r^{2}/C^{2}}{(1+2r^{2}/A^{2})^{2}}\right]
\label{tol-iv-rho}\\
8\pi P  &  =\frac{1}{A^{2}}\left[  \frac{1-(A^{2}/C^{2}+3r^{2}/C^{2}%
)}{1+2r^{2}/A^{2}}\right]  \label{tol-iv-p}%
\end{align}
Constants $A$ and $C$ can be expressed in terms of the central fluid values.
We have
\begin{align*}
8\pi\rho_{c}  &  =\frac{3}{A^{2}}\left[  1+A^{2}/C^{2}\right] \\
8\pi P_{c}  &  =\frac{1}{A^{2}}\left[  1-A^{2}/C^{2}\right] \\
A^{2}  &  =\frac{2}{8\pi(\rho_{c}/3+P_{c})}\\
C^{2}  &  =\frac{2}{8\pi(\rho_{c}/3-P_{c})}%
\end{align*}
Note that the central fluid equation of state (EOS) is constrained:
$P_{c}<\rho_{c}/3.$ The zero-pressure boundary that occurs in the regular
derivative match has size%
\begin{equation}
R_{z}^{2}=C^{2}/3-A^{2}/3.
\end{equation}

\subsubsection{Metric Match}

The match to vacuum Schwarzschild provides
\begin{equation}
B^{2}(1+R_{0}^{2}/A^{2})=1-2m_{0}/R_{0}.
\end{equation}
The fractional match is%
\begin{align}
\frac{B^{2}}{A^{2}}\frac{1}{R_{0}^{\alpha-2}}\frac{1}{\Gamma(3-\alpha)}  &
=\frac{m_{0}}{R_{0}^{1+\alpha}}\Gamma(1+\alpha)\\
B^{2}R_{0}^{3}  &  =A^{2}m_{0}\ \Gamma(3-\alpha)\Gamma(1+\alpha)\nonumber
\end{align}
Combining with the metric match we obtain%
\begin{align}
A^{2}  &  =R_{0}^{2}\frac{(R_{0}/m_{0})-[2+\Gamma(3-\alpha)\Gamma(1+\alpha
)]}{\Gamma(3-\alpha)\Gamma(1+\alpha)}\\
B^{2}  &  =\frac{m_{0}}{R_{0}}\left[  \frac{R_{0}}{m_{0}}-[2+\Gamma
(3-\alpha)\Gamma(1+\alpha)]\right] \nonumber\\
R_{0}/m_{0}  &  >2+\Gamma(3-\alpha)\Gamma(1+\alpha)\nonumber
\end{align}
The boundary size depends on the central EOS as well as the order of the
fractional derivative.%
\begin{equation}
(\frac{R_{0}}{m_{0}})^{3}-(\frac{R_{0}}{m_{0}})^{2}[2+\Gamma(3-\alpha
)\Gamma(1+\alpha)]-\frac{\Gamma(3-\alpha)\Gamma(1+\alpha)}{4\pi m_{0}%
^{2}(P_{c}+\rho_{c}/3)}=0
\end{equation}

\subsubsection{Crust Stress-Energy}

We introduce scaled parameters $r_{A}^{2}:=R_{0}^{2}/A^{2},$ $r_{C}^{2}%
:=R_{0}^{2}/C^{2},$ and $r_{z}^{2}:=R_{z}^{2}/A^{2}$.%
\begin{align}
8\pi S_{0}^{0}  &  =\frac{1}{(g_{yy}^{E})^{1/2}}\frac{g_{\theta\theta,y}^{E}%
}{g_{\theta\theta}^{E}}-\frac{1}{(g_{rr}^{I})^{1/2}}\frac{g_{\theta\theta
,r}^{I}}{g_{\theta\theta}^{I}}\nonumber\\
&  =\frac{2}{R_{0}}\left[  \sqrt{1-2m_{0}/R_{0}}-\sqrt{\frac{(1-r_{C}%
^{2})(1+r_{A}^{2})}{1+2r_{A}^{2}}}\right] \nonumber\\
&  =\frac{2}{R_{0}}\sqrt{1+r_{A}^{2}}\left[  B-\frac{A}{C}\sqrt{\frac
{1+3r_{z}^{2}-r_{A}^{2}}{1+2r_{A}^{2}}}\right] \\
8\pi S_{\theta}^{\theta}  &  =8\pi S_{\phi}^{\phi}=\frac{1}{2}\left[  \frac
{1}{(g_{yy}^{E})^{1/2}}\frac{g_{\theta\theta,y}^{E}}{g_{\theta\theta}^{E}%
}-\frac{1}{(g_{rr}^{I})^{1/2}}\frac{g_{\theta\theta,r}^{I}}{g_{\theta\theta
}^{I}}\right]  +\frac{1}{2}\left[  \frac{1}{(g_{yy}^{E})^{1/2}}\frac
{g_{00,y}^{E}}{g_{00}^{E}}-\frac{1}{(g_{rr}^{I})^{1/2}}\frac{g_{00,r}^{I}%
}{g_{00}^{I}}\right] \nonumber\\
&  =\frac{1}{R_{0}}\sqrt{1+r_{A}^{2}}\left[  B-\frac{A}{C}\sqrt{\frac
{1+3r_{z}^{2}-r_{A}^{2}}{1+2r_{A}^{2}}}\right] \\
&  +\frac{1}{R_{0}\sqrt{1+r_{A}^{2}}}\left[  \frac{m_{0}/R_{0}}{B}-(r_{A}%
^{2})\sqrt{\frac{1-r_{C}^{2}}{1+2r_{A}^{2}}}\right] \nonumber
\end{align}

Some examples of radius variation and crust stress energy are given in Section V.

\section{Equilibrium in the Presence of Surface Stresses}

\subsection{Stress-Energy}

The Israel layer is the zero-thickness idealization of a bounding layer with
finite thickness, $d$. The physical crust runs from an outer boundary $R_{0}$
to an interior fluid boundary $R_{i}$ with $d=R_{0}-R_{i.}$ We know the
interior fluid solutions will satisfy the \cite{Wal84}
Tolman-Oppenheimer-Volkov (TOV) equation. The Israel layers generated in this
work are obtained by introducing a discontinuity in the derivative of $g_{00}%
$.\ The analog of the TOV equation for the layer, requiring that the solutions
remain static, will provide relations among the model parameters.\ To develop
the TOV analog for the layer, consider the general static spherical metric for
an interior fluid with pressure $P$ and density $\rho$
\begin{equation}
ds^{2}=-e^{\nu}dt^{2}+e^{\lambda}dr^{2}+r^{2}d\Omega^{2}. \label{gen-stat-met}%
\end{equation}
The details of the field equations are given in Appendix B.\ The covariant
derivative of the general energy-momentum tensor provides the conservation
equation
\begin{equation}
-\frac{\partial T_{r}^{r}}{\partial r}-(\frac{\nu^{\prime}}{2}+\frac{2}%
{r})T_{r}^{r}+(\frac{\nu^{\prime}}{2})T_{0}^{0}+(\frac{2}{r})T_{\theta
}^{\theta}=0 \label{div-en-mom}%
\end{equation}
For an isotropic fluid matched to vacuum, this is the usual TOV equation%
\begin{equation}
\frac{\partial P}{\partial r}+\frac{\nu^{\prime}}{2}(P+\rho)=0.
\label{tov-eqn}%
\end{equation}
It is the analog of this equation which we want for the Israel layer.

\subsection{The conservation equation over a limiting shell}

Consider a bounding shell which will approximate a thin surface layer.\ The
central radius of the shell is $R$ with the outer boundary $R^{(+)}=R+d/2$,
and the inner interior fluid boundary at $R^{(-)}=R-d/2$.\ $d$ is the
coordinate shell thickness. In the $d\rightarrow0$ limit, $R\rightarrow
R_{0}.$ A general stress-energy $T_{j}^{i}$ can be related to a surface
stress-energy $S_{b}^{a}$ by \cite{Poi04}
\begin{equation}
T_{j}^{i}=\delta(l)S_{b}^{a}e_{a}^{i}e_{j}^{b}%
\end{equation}
where $e_{a}^{i}$ is a tangent vector to the shell, and $l$ the proper
distance along a radial geodesic, $l=e^{\lambda/2}d$. \ The shell
stress-energy has a perfect fluid analogue%
\begin{align*}
S^{ij}  &  =\sigma U^{i}U^{j}+\tau(h^{ij}+U^{i}U^{j})\\
h^{ij}  &  =g^{ij}-n^{i}n^{j}\\
n^{i}  &  =(0,e^{-\lambda},0,0)
\end{align*}
where $S_{0}^{0}/c^{2}=-\sigma($gm/cm$^{2})$ and $S_{\theta}^{\theta}=\pm
\tau($dynes/cm$).$ Following Poisson \cite{Poi04}, we take $l=0$ on the
hypersurface defined by $R$, with $l$ negative for $r<R$ and positive on the
vacuum side, $r>R$. \ The $T_{r}^{r}$ content of the shell can be described
using a Heaviside function, $\Theta(l),$ as%
\begin{equation}
T_{r}^{r}=\Theta(l)T_{r}^{(+)r}+\Theta(-l)T_{r}^{(-)r}+\delta(l)S_{r}^{r}.
\end{equation}
The last term will be zero for the $2+1$ shell stress energy.\ Forming the
derivative needed in the conservation equation we have%
\[
\frac{\partial T_{r}^{r}\ }{\partial r}=\delta(l)\frac{dl}{dr}T_{r}%
^{(+)r}+\Theta(l)\frac{\partial T_{r}^{(+)r}}{\partial r}-\delta(l)\frac
{dl}{dr}T_{r}^{(-)r}+\Theta(-l)\frac{\partial T_{r}^{(-)r}}{\partial r}%
\]
In the $l\rightarrow0$ limit we have%
\begin{equation}
\frac{\partial T_{r}^{r}}{\partial r}=-\lim_{l\rightarrow0}[\delta(l)\frac
{dl}{dr}T_{r}^{(-)r}]=-\lim_{l\rightarrow0}[\delta(l)Pe^{\lambda/2}]
\end{equation}
where the first term is zero with no radial pressure on the outer boundary.
The stress-energy function evaluated at the inner boundary is $P$ and
$\lambda^{(-)}$ is an interior metric function. Substituting into the
conservation equation in the $l\rightarrow0$ hypersurface limit we have%
\begin{equation}
Pe^{\lambda^{(-)}/2}+(\nu^{\prime}/2)\ S_{0}^{0}+(2/R_{0})\ S_{\theta}%
^{\theta}=0. \label{p-nu-prime}%
\end{equation}

\subsection{Evaluating $\partial_{r}$g$_{00}$}

The derivative, $\nu^{\prime},$ on the hypersurface can be written as a
difference equation%
\begin{align*}
\nu^{\prime}(R)  &  =\frac{\nu(R+d/2)-\nu(R-d/2)}{(R+d/2)-(R-d/2)}=\frac
{\nu(R+d/2)-\nu(R)+\nu(R)-\nu(R-d/2)}{(R+d/2)-(R-d/2)}\\
&  =\frac{\nu(R+d/2)-\nu(R)}{d}+\frac{\nu(R)-\nu(R-d/2)}{d}%
\end{align*}
Expanding, we can write%
\[
\nu(R\pm d/2)=\nu(R)\pm\nu^{\prime}[R^{(\pm)}]\frac{d}{2}+...
\]
Substituting in the thin shell limit we have%
\[
\nu^{\prime}(R)\approx\frac{\nu^{\prime}[R^{(+)}]+\nu^{\prime}[R^{(-)}]}{2}%
\]
The first term follows from the Schwarzschild metric match and the second is
given in Appendix B.\ We have%
\[
\nu^{\prime}(R)\approx(2m/R^{2}+4\pi RP)(1-2m/R)^{-1}%
\]
where we have identified the Schwarzschild mass parameter with the interior
mass of the fluid. Substituting into Eq.(\ref{p-nu-prime}) we have%
\begin{equation}
-P(1-2m/R)^{-1/2}=(m/R^{2}+2\pi RP)(1-2m/R)^{-1}S_{0}^{0}+(2/R)S_{\theta
}^{\theta}%
\end{equation}
which describes the thin shell pressure balance. The classical limit of this
equation follows from $c\rightarrow\infty$ and is%
\[
P-\sigma\frac{m}{R^{2}}=(\frac{2}{R})(-\tau).
\]
If the fluid pressure at the interior boundary dominates, this can be
interpreted as a tension in the shell balancing the outward interior fluid
pressure at the boundary minus the inward pressure due to the gravitational
attraction of the shell by the interior fluid. If the shell mass term
dominates, the stress in the boundary layer will be a pressure. In the next
section we explore the stress-energy structure of the boundary layer and will
see parameter ranges with both layer tension and pressure.

\section{Details of the Fractional Match}

\subsection{Sphere Radii}

The sizes of the sphere are described by%
\begin{align}
\text{Tolman V} &  :\frac{R_{0}}{2m_{0}}=1+\frac{\Gamma(1+\alpha
)\Gamma(4n/(1+n)+1-\alpha)}{\Gamma(1+4n/(1+n))}\\
\text{Tolman IV} &  :(\frac{R_{0}}{m_{0}})^{3}-(\frac{R_{0}}{m_{0}}%
)^{2}[2+\Gamma(3-\alpha)\Gamma(1+\alpha)]-\frac{\Gamma(3-\alpha)\Gamma
(1+\alpha)}{4\pi m_{0}^{2}(P_{c}+\rho_{c}/3)}=0\label{cubic}%
\end{align}
The scaled boundary radius, $R_{0}/m_{0}$, for Tolman V is plotted as a
function of alpha for various $n$ in Figure 1. \
\begin{figure}
[ptb]
\begin{center}
\includegraphics[
height=3.5985in,
width=4.7755in
]%
{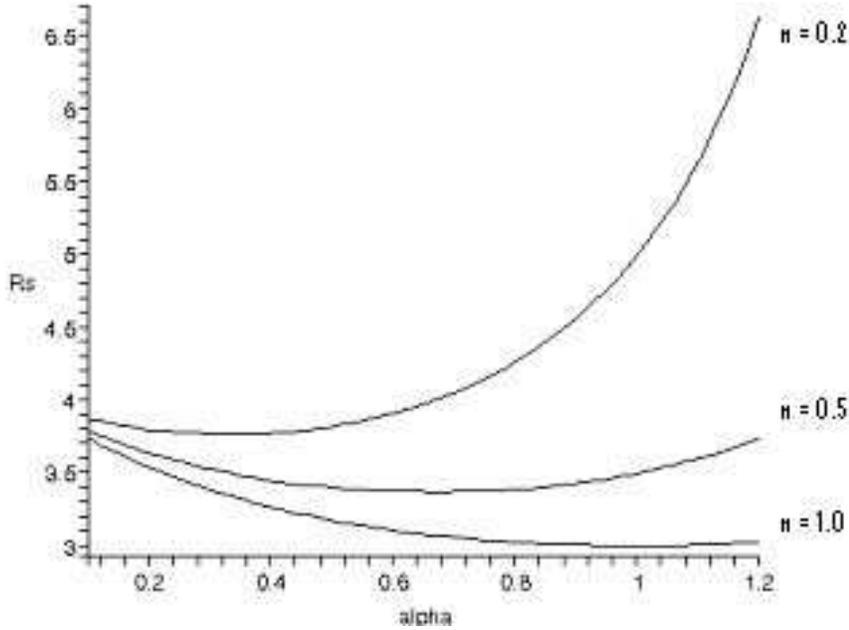}%
\caption{Scaled radius vs fractional order}%
\end{center}
\end{figure}
\ \ The overall effect of the fractional match is to increase the range of
sphere sizes for a given EOS. The largest differences are for low $n$ fluids
where a very much smaller sphere radius is possible than for the zero-pressure
match. \ The Buchdahl bound \cite{Buc59} limits the ratio of $2m_{0}/R_{0}$
for fluid spheres whose $g_{00}$ component is continuous across the boundary
and whose density is decreasing outward. We have matched fractional
derivatives rather than first derivatives and it is not clear that the
conditions of the Buchdahl bound are satisfied, but\ from Figure 1, it is seen
that the Buchdahl bound, $2m_{0}/R_{0}\leq8/9,$ is not violated.

The radius for Tolman IV is a cubic root of Eq.(\ref{cubic}) but some general
description can be given. \ The modeling term in the equation is the
denominator of the last term.\ Consider the factor
\[
c_{1}\sim4\pi m_{0}^{2}\ \rho_{c}/3
\]
describing an object with mass $m_{0}=Nm_{\bigodot}=N(2\times10^{30}kg)$, and
central density and pressure of neutron star order, $\rho_{c}\sim
10^{17}kg/m^{3}$, $P_{c}\sim10^{33}Newton/m^{2}$.\textbf{ }Numerical scaled
radius values using these values are described in Table I for a range of N and
alpha values. For $N\sim100$ or larger, the last term in the cubic is
negligible and the radius is essentially given by the limiting value%
\[
R_{0}/m_{0}\sim2+\Gamma(3-\alpha)\Gamma(1+\alpha)
\]
The masses for these radii are well out of the neutron star range. The
reflection symmetry about $\alpha=1$ is the result of a product equivalence of
the two gamma functions for paired alpha values, i.e. \ $\alpha=(0.6,1.4)$
give the same gamma function product. From Table I it is clear that the low
$N$ values have masses and radii of neutron star orders of magnitude
\cite{BL05}. For example, for $N=1$, the radii are approximately $R_{0}%
\sim11.44m_{0}\sim17km$ . Smaller central densities, describing more ordinary
fluid objects, result in much larger fluid spheres. For a central density of
\ $\rho_{c}\sim10^{10}kg/m^{3},$ the radii for $\alpha=1,$ $R_{0}(N)$ are
\ $R_{0}(1)=2228.33m_{0},$ $R_{0}(10)=480.87m_{0},$ $R_{0}(100)=104.39m_{0},$
$R_{0}(1000)=23.32m_{0}$, with the values for larger and smaller $\alpha$
paired and increasing, just as for the larger central density.

\ \ \ \ \ \ \ \ \ \ %

\begin{tabular}
[c]{|c|c|c|c|c|}\hline
$\alpha$ & $N=1$ & $N=10$ & $N=100$ & $N=1000$\\\hline
$0.2$ & $13.24$ & $4.41$ & $3.55$ & $3.54$\\\hline
$0.4$ & $12.39$ & $4.10$ & $3.28$ & $3.27$\\\hline
$0.6$ & $11.85$ & $3.91$ & $3.12$ & $3.11$\\\hline
$0.8$ & $11.54$ & $3.81$ & $3.04$ & $3.03$\\\hline
$1$ & $11.44$ & $3.78$ & $3.01$ & $3$\\\hline
$1.2$ & $11.54$ & $3.81$ & $3.04$ & $3.03$\\\hline
$1.4$ & $11.85$ & $3.91$ & $3.12$ & $3.11$\\\hline
$1.6$ & $12.39$ & $4.10.$ & $3.28$ & $3.27$\\\hline
\end{tabular}

~\newline Table I: \ $R_{0}/M_{0}$ - Tolman IV-$\rho_{c}\sim10^{17}kg/m^{3}$

\subsection{Crust Stress-Energy}%

\begin{figure}
[h]
\begin{center}
\includegraphics[
height=3.5985in,
width=5.9248in
]%
{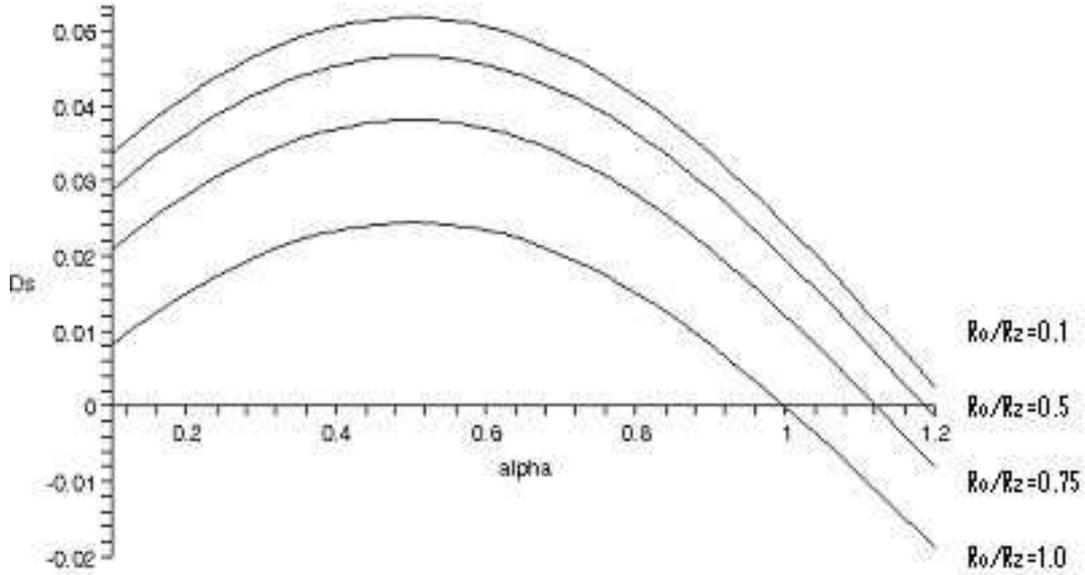}%
\caption{$Ds=-8\pi m_{0}S_{0}^{0}$ vs fractional order}%
\end{center}
\end{figure}
\begin{figure}
[h]
\begin{center}
\includegraphics[
height=3.5942in,
width=6.6755in
]%
{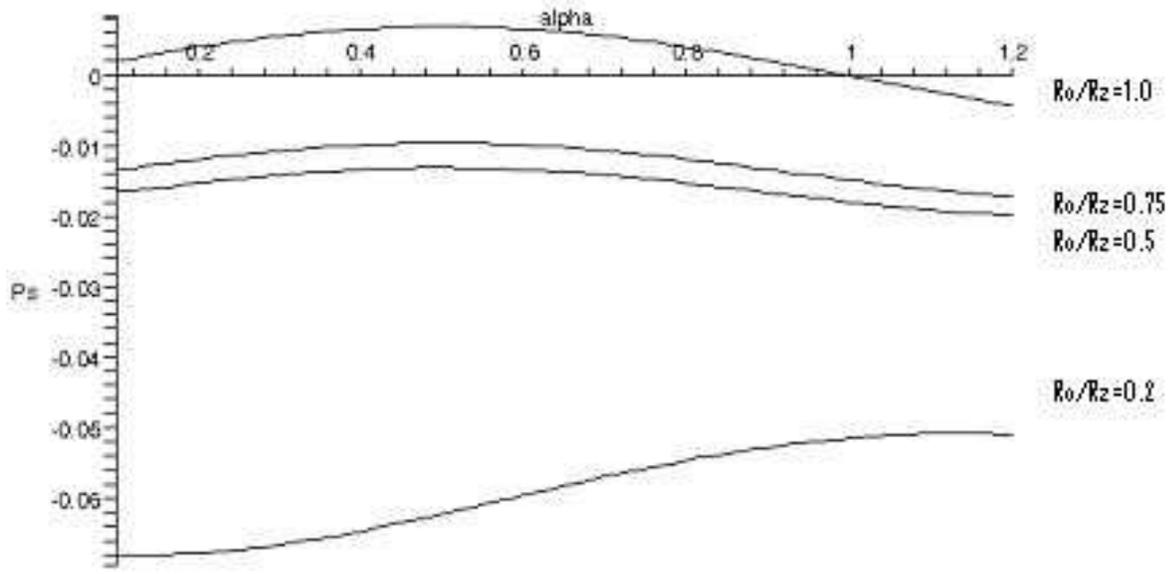}%
\caption{$Ps=8\pi m_{0}S_{\vartheta}^{\vartheta}$ vs fractional order}%
\end{center}
\end{figure}
Figures $2$ and $3$ describe the variation of the boundary energy density and
pressure, in Tolman V, $n=1/3$, as the size of the fluid sphere varies.\ Over
a large part of the $R_{0}/R_{z}$ range, a crust tension contains the interior
fluid, with the tension increasing in size as the fluid sphere becomes
smaller, essentially acting to squeeze the fluid into smaller volumes. The
results are similar for $C=0$.

For Tolman IV, the crust energy density is%
\begin{align*}
8\pi\sigma &  =\frac{2}{R_{0}}\left[  -\sqrt{1-2m_{0}/R_{0}}+\sqrt
{\frac{(1-R_{0}^{2}/C^{2})(1+R_{0}^{2}/A^{2})}{1+2R_{0}^{2}/A^{2}}}\right] \\
&  =\frac{2}{R_{0}}\left[  -\sqrt{1-2m_{0}/R_{0}}+\sqrt{\frac{[1-4\pi
R_{0}^{2}\rho_{c}(1/3-n)][(1+4\pi\rho_{c}R_{0}^{2}(1/3+n)]}{1+R_{0}^{2}%
8\pi\rho_{c}(1/3+n)}}\right]
\end{align*}
The modeling factor of importance is the term
\[
4\pi\frac{R_{0}^{2}}{m_{0}^{2}}\rho_{c}m_{0}^{2}=3c_{1}\frac{R_{0}^{2}}%
{m_{0}^{2}}%
\]
For nuclear central densities, $\rho_{c}\sim10^{14}g/cm^{3}$, this is
\[
20.35N^{2}\times10^{-4}\frac{R_{0}^{2}}{m_{0}^{2}}.
\]
In order to have real values we require
\begin{align*}
N  &  =1,\text{ \ }R_{0}\sim12m_{0},\text{ }[0.293\times(1/3-n)]<1\\
N  &  =10,\text{ }R_{0}\sim4m_{0},\text{ }[3.26\times(1/3-n)]<1\\
N  &  =100,\text{ }R_{0}\sim3m_{0},\text{ }[183\times(1/3-n)]<1
\end{align*}
It is clear that the broadest range of\ central equations of state for nuclear
central densities is for the lower mass objects.$\ $\ Higher mass objects
require a central EOS very close to the $1/3$ limit. \ For smaller values of
the central density the central EOS range is much broader. For a central
density of \ $\rho_{c}\sim10^{7}g/cm^{3},$ the modeling factor is
\[
4\pi\frac{R_{0}^{2}}{m_{0}^{2}}\rho_{c}m_{0}^{2}=20.35N^{2}\times10^{-11}%
\frac{R_{0}^{2}}{m_{0}^{2}}%
\]
and for real values require%
\begin{align*}
N  &  =1,\text{ \ \ \ \ }R_{0}\sim2228m_{0},\text{ }[0.10\ \times
10^{-2}(1/3-n)]<1\\
N  &  =10,\text{ \ \ \ }R_{0}\sim480m_{0},\text{ }[0.47\times10^{-2}%
(1/3-n)]<1\\
N  &  =100,\text{ \ \ }R_{0}\sim104m_{0},\text{ }[0.022\times(1/3-n)]<1\\
N  &  =1000,\text{ }R_{0}\sim23m_{0},\text{ }[0.11\times(1/3-n)]<1
\end{align*}

\section{Conclusion}

In this work we have examined a family of boundary layers created by matching
fractional derivatives across a boundary. The boundary layers considered have
structure which depends on the order of the fractional derivative. One of the
reasons that fractional calculus may be important for boundary layers is the
mechanism by which a boundary layer is formed.\ One of the possible ways to
build a variable density crust is by a diffusive process; a process whose
underlying cause is Brownian motion.\textbf{ }This motion, as analyzed by
statistical mechanics, involves diffusion, dissipation, and the
fluctuation-dissipation theorem. The dynamical model of Brownian motion was
provided by Langevin in 1908 using a stochastic differential equation. It
seems apparent from the nature of randomness that such macroscopic stochastic
equations are incompatible with the continuous and differentiable character of
microscopic Hamiltonian dynamics. (Think of the conventional diffusion
equation, with the diffusion process described by a second order spatial
derivative.) Therefore, the mathematical description rests on either ordinary
analytical functions describing the dynamics, or on conventional differential
operators describing the phase space evolution. The differentiable nature of
the macroscopic picture is, in a sense, a natural consequence of microscopic
randomness. This means that use can be made of ordinary differential
calculations on the macroscopic scale, even if the microscopic dynamics are
incompatible with ordinary calculus methods. On the other hand, in the case
where a timescale separation between macroscopic and microscopic levels of
description does not exist, the non-differentiable nature of the microscopic
dynamics is transmitted to the macroscopic level.\textbf{\ }Since fractional
calculus has been shown to provide a good description for a range of diffusive
processes \cite{Pod99}, one might expect a boundary condition based on
fractional calculus would reflect the fractional growth process.\ An example,
given by Allegrini, Grigolini, and West \cite{AGW96}, shows that a diffusion
process generated by a fluctuation with no time scale at the macroscopic level
generates a diffusion process well described by a fractional Laplacean.

While diffusion is a possible mechanism for generating a layer with structure,
the method of generating the fractional family of layers is independent of the
production mechanism and generates a family whose stress-energy and size
depends on the order of the fractional derivative. The result is a much
broader range in fluid sphere properties. For example, in Tolman V, a much
larger range of spheres sizes can be described with the fractional layer than
without, with the energy density of the layer decreasing as the size of the
sphere increases.\ The $C\neq0$ Tolman V spheres have a zero-pressure boundary
solution.\ For spheres smaller than the zero-pressure sphere, the layer has a
tension, while for spheres larger than the zero-pressure sphere, the layer has
positive stress over much of the range of the fractional order.\ The
fractional boundary could prove to be a valuable modeling tool in more
realistic neutron star models.

The range of stress-energys in the fractional boundary layers implies
differences in structure as a function of the fractional order. The
differences in density could be modeled in several ways: for example, with
different crust materials or different incomplete fluid coverings (tilings)
\cite{Kri05}. The layer itself is a model of a thin crust and there could be
differences in the geometry of the 2+1 shells that fill the real crust.\ The
interior fluid geometry and the exterior Schwarzschild vacuum do not have to
match for a crust with finite thickness.

The models presented in this paper matched an integral transform of the
regular derivative across a spatial boundary. It is not a fractional
generalization of general relativity, but a fractional genralization of a
boundary condition. The next step is to explore the range of fractional
generalizations of other sets of boundary conditions\ and other derivative
definitions to applicable spacetimes.\textbf{ }

\appendix

\section{DERIVATIVES}

\subsection{Regular}

For functions $f$ and $F$ continuous on $[a,b]$ $\epsilon$ \textit{Reals}%
\[
F(x):=%
{\textstyle\int\limits_{a}^{x}}
f(t)dt
\]
$F(x)$ is differentiable such that $dF/dx=f$. The $n^{th}$ integer derivative
is simply $d^{n}F/dx^{n}$. For example
\[
\frac{d^{n}}{dx^{n}}x^{k}=\frac{k!}{(k-n)!}x^{k-n}.
\]
With the gamma function this is
\[
\frac{d^{n}}{dx^{n}}x^{\alpha}=\frac{\Gamma(\alpha+1)}{\Gamma(\alpha
-n+1)}x^{\alpha-n}.
\]
The gamma function, $\Gamma(z)$, is defined as%
\begin{align*}
\Gamma(z)  &  =%
{\textstyle\int\limits_{0}^{\infty}}
e^{-t}t^{z-1}dt,\text{ \ }\Gamma(1/2)=\sqrt{\pi},\text{ \ }\Gamma(1)=1\\
\Gamma(n+1)  &  =n\Gamma(n)=n!\text{ \ for }n>0.
\end{align*}

\subsection{Fractional}

\subsubsection{Riemann-Liouville}

The Riemann-Liouville definition for the $\alpha$ fractional derivative of
$f(x)$ is, with $\alpha\geq0$,%
\begin{equation}
D^{\alpha}f(x)=\frac{d^{\alpha}}{dx^{\alpha}}f(x):=\frac{1}{\Gamma(n-\alpha
)}\frac{d^{n}}{dx^{n}}%
{\textstyle\int\limits_{c}^{x}}
\frac{f(t)}{(x-t)^{\alpha-n+1}}dt \label{gen-r-l}%
\end{equation}
where $n$ is the smallest integer larger than $\alpha$ when it is fractional,
that is $n=[\alpha]+1$. In the $\alpha=1$ limit, the derivative produces the
integer result. The constant $c$ in the limit of the integral is usually set
to $0$ (Riemann definition) or to $-\infty$ (Liouville definition). For
example, the Riemann-Liouville derivative of $x^{k}$ for $\alpha\leq1$ with
$n=1$ we have%
\begin{align*}
D^{\alpha}x^{k}  &  =\frac{1}{\Gamma(1-\alpha)}\frac{d}{dx}%
{\textstyle\int\limits_{0}^{x}}
t^{k}(x-t)^{-\alpha}\text{ }dt\\
&  =\frac{1}{\Gamma(1-\alpha)}\frac{d}{dx}%
{\textstyle\int\limits_{0}^{x}}
t^{k}x^{-\alpha}(1-\frac{t}{x})^{-\alpha}\text{ }dt\\
&  =\frac{1}{\Gamma(1-\alpha)}\frac{d}{dx}%
{\textstyle\int\limits_{0}^{1}}
w^{k}x^{-\alpha+1+k}(1-w)^{-\alpha}\text{ }dw
\end{align*}
Using the definition of the beta function%
\[%
{\displaystyle\int\limits_{0}^{1}}
w^{p-1}(1-w)^{q-1}dw=\frac{\Gamma(p)\Gamma(q)}{\Gamma(p+q)}%
\]
we have%
\begin{align*}
D^{\alpha}x^{k}  &  =\frac{1}{\Gamma(1-\alpha)}\frac{dx^{-\alpha+1+k}}%
{dx}\frac{\Gamma(k+1)\Gamma(1-\alpha)}{\Gamma(k+2-\alpha)}\\
&  =(1+k-\alpha)x^{k-\alpha}\frac{\Gamma(k+1)}{\Gamma(k+2-\alpha)}\\
&  =x^{k-\alpha}\frac{\Gamma(k+1)}{\Gamma(k+1-\alpha)}%
\end{align*}
For $\alpha=1,$ this is the usual result $D^{1}x^{k}=dx^{k-1}.$ Note that, for
$k=-1$, this operation fails.\ This is an example of one of the problems
encountered in applying fractional derivatives to general relativity. Not all
definitions of fractional derivatives work for all functions.\ The fractional
derivative of $x^{k}$ for $\alpha\geq1$ is identical to the fractional
derivative for $\alpha\leq1.$ For this function, the derivative is continuous
across the $\alpha=1$ boundary. \
\begin{align*}
D^{\alpha}x^{k}  &  =\frac{1}{\Gamma(2-\alpha)}\frac{d^{2}}{dx^{2}}%
{\textstyle\int\limits_{0}^{x}}
t^{k}x^{1-\alpha}(1-\frac{t}{x})^{1-\alpha}\text{ }dt\\
&  =\frac{1}{\Gamma(2-\alpha)}\frac{d^{2}}{dx^{2}}%
{\textstyle\int\limits_{0}^{1}}
w^{k}x^{k+2-\alpha\ }(1-w)^{1-\alpha}\text{ }dw\\
&  =x^{k-\alpha}\frac{\Gamma(1+k)}{\Gamma(1+k-\alpha)}%
\end{align*}
One should note that the Riemann-Liouville fractional derivative of a constant
is not zero. \ 

\subsubsection{Caputo}

The Caputo derivative is the integral transform of the regular derivative and
is found by moving the derivative in the Riemann-Liouville definition inside
the integral to act on the function. We have%
\begin{equation}
D^{\alpha}f(x)=\frac{1}{\Gamma(n-\alpha)}%
{\textstyle\int\limits_{0}^{x}}
\frac{\frac{d^{n}}{dt^{n}}f(t)}{(x-t)^{\alpha-n+1}}dt \label{gen-caputo}%
\end{equation}
Example: $f(x)=x^{b}$, $\beta\geq0$%
\[
D^{\alpha}(x-a)^{k}=\frac{1}{\Gamma(n-\alpha)}%
{\textstyle\int\limits_{0}^{x}}
\frac{\frac{d^{n}}{dt^{n}}t^{k}}{(x-t)^{\alpha-n+1}}dt
\]
For $\alpha\leq1,$ $n=1$ we have%
\begin{align}
D^{\alpha}(x-a)^{k}  &  =\frac{1}{\Gamma(1-\alpha)}%
{\textstyle\int\limits_{0}^{x}}
kt^{k-1}(x-t)^{-\alpha}dt\nonumber\\
&  =\frac{kx^{k-\alpha}}{\Gamma(1-\alpha)}%
{\textstyle\int\limits_{0}^{1}}
w^{k-1}(1-w)^{-\alpha}dw\nonumber\\
&  =kx^{k-\alpha}\frac{\Gamma(k)}{\Gamma(k+1-\alpha)} \label{caputo-r-l}%
\end{align}
which is identical to the Riemann-Liouville derivative for this function. This
derivative also is not defined for for $k=-1$. In general relativity, one of
the spacetimes one would like to treat is vacuum Schwarzschild but the
Riemann-Liouville derivative will not give finite answers for the $1/r$
structure. \ The derivative of $1/r$ can be taken with the Weyl derivative.

\subsubsection{Weyl}

The Weyl derivative differs from the Riemann-Liouville derivatives over the
range of the fractional transform. To take the fractional derivatives of $1/r$
we use the Weyl derivative over the range $\ (R_{0}$,$\infty).$ The Weyl
derivative of $f(r)$ can be written as%
\[
D^{\alpha}f(r)=\frac{(-1)^{n-1}}{\Gamma(n-\alpha)}%
{\displaystyle\int\limits_{r}^{\infty}}
\ \frac{d^{n}f(t)}{dt^{n}}(t-r)^{n-\alpha\ }dt
\]
where $n$ is the smallest integer above $\alpha$ when it is fractional. This
paper is concerned with the fractional derivative across the boundary and the
phase $(-1)^{n-1},$ was chosen to make $D^{\alpha}$ continuous across
$\alpha=1.$ Applying the derivative definition to $1/r$ for $\alpha\leq1$
$(n=1)$ we find%
\begin{equation}
D^{\alpha}r^{-1}=-r^{-(1+\alpha)}\Gamma(1+\alpha),\text{ \ \ \ }\alpha\leq1
\label{1/r-deriv}%
\end{equation}
For $\alpha=1,$ this gives the usual first derivative of$\ 1/r.$ For
$\alpha>1$ $(n=2)$ the derivative is the same. One should be careful not to
interpret the derivative for $\alpha=2$ as the second derivative. The second
derivative would follow from a double application of $D^{\alpha}$.\ For this
function the single derivative at $\alpha=2$ is not the same as the double
application of the derivative operator.

\section{Fluid Sphere Formalism}

Consider the general static spherical metric over the interior fluid with
$\nu(r)$ and $\lambda(r)$%
\begin{equation}
ds^{2}=-e^{\nu}dt^{2}+e^{\lambda}dr^{2}+r^{2}d\Omega^{2}.
\end{equation}
With Einstein's field equations as $G_{ij}=8\pi T_{ij}$, the energy-momentum
components are%
\begin{align*}
8\pi T_{\text{ }0}^{0}  &  =-e^{-\lambda}(\lambda^{\prime}/r-1/r^{2}%
)-1/r^{2}\\
8\pi T_{\text{ }r}^{r}  &  =e^{-\lambda}(\nu^{\prime}/r+1/r^{2})-1/r^{2}\\
8\pi T_{\text{ }\theta}^{\theta}  &  =8\pi T_{\phi}^{\phi}=(e^{-\lambda
}/2)[\nu^{\prime\prime}+(\nu^{\prime}/2+1/r)(\nu^{\prime}-\lambda^{\prime})]
\end{align*}
For the fluid interior, the energy-momentum, with four-velocity $U^{i}%
=(e^{\nu/2},0,0,0)$, is%
\[
T^{ij}=(\rho+P)U^{i}U^{j}+Pg^{ij}.
\]
In the comoving frame the fluid stress energy is%
\[
T_{\text{ }0}^{0}=-\rho,\text{ \ }T_{\text{ }r}^{r}=T_{\text{ }\theta}%
^{\theta}=T_{\text{ }\phi}^{\phi}=P.
\]
It is common to use the function $m(r)$ in $g_{rr}$ with%
\[
e^{\lambda}=[1-2m(r)/r]^{-1},
\]
so that%
\[
\lambda^{\prime}=2(m^{\prime}/r-m/r^{2})(1-2m/r)^{-1}.
\]
From the first field equation we have$\ $%
\[
m^{\prime}=4\pi r^{2}\rho
\]
The second field equation provides a relation between the fluid pressure, $P$,
and $\nu^{\prime}$%
\[
\nu^{\prime}/2=(4\pi rP+m/r^{2})(1-2m/r)^{-1}%
\]

\end{document}